\documentclass[aps,prb,reprint,superscriptaddress,longbibliography,floatfix]{revtex4-2}
\usepackage[utf8]{inputenc}
\usepackage[T1]{fontenc}
\usepackage{graphicx}
\usepackage{amsmath}
\usepackage{amssymb}
\usepackage{bm}
\usepackage{xcolor}
\usepackage[version=4]{mhchem}

\begin{document}

\title{Programmable Photocatalysis via Symmetry-Defined Periodic Potentials}

\author{Qun Yang}
\email{qun.yang@mpi-halle.mpg.de}
\affiliation{Max Planck Institute of Microstructure Physics, Halle (Saale) 06120, Germany}
\affiliation{Division of Physical Sciences, College of Letters and Science, University of California, Los Angeles, California 90095, USA}

\author{Di Luo}
\affiliation{Department of Electrical and Computer Engineering, University of California, Los Angeles, California 90095, USA}

\author{Prineha Narang}
\affiliation{Division of Physical Sciences, College of Letters and Science, University of California, Los Angeles, California 90095, USA}

\date{\today}

\begin{abstract}
Photocatalysis in atomically thin semiconductors is often limited by rapid electron-hole recombination, making it difficult to translate favorable band structures into efficient chemical function. Here we propose symmetry-defined periodic potentials as a strategy for photocatalysis: instead of modifying the chemistry of the active layer, one engineers a long-wavelength electrostatic landscape that spatially separates photoexcited electrons and holes. Applied to monolayer InSe, we show that experimentally accessible moir\'e patterns, such as those generated by twisted hBN, produce miniband formation, band-gap renormalization, and robust carrier separation. Using commensurate BN/InSe local registries, we further show that the moir\'e control layer transfers a measurable electrostatic modulation to InSe, providing the microscopic link between continuum potential engineering and the local surface environment. The key result is that the periodic potential strongly reorganizes carrier distribution while only weakly perturbing adsorption trends, thereby identifying a practically useful regime in which charge separation can be engineered without demanding major changes to the underlying surface chemistry. These results position periodic potentials as a broadly applicable design principle for photocatalysis and other light-driven interfacial phenomena in two-dimensional materials.
\end{abstract}

\maketitle

%\section{Introduction}

%\textcolor{blue}{Controlling the spatial distribution and dynamics of electronic states in low-dimensional materials is a central challenge in condensed matter physics and materials science, with direct implications for optoelectronics, exciton physics, and interfacial charge transfer processes\cite{menke2018order,rahman2022hole}} \textcolor{blue}{In particular, the generation, separation, and transport of photoexcited carriers play a critical role in applications such as photocatalysis, photovoltaics, and light-harvesting systems\cite{zhang2025charge}.} 
Photocatalysis offers a sustainable approach to harnessing solar energy for chemical transformations, enabling applications such as water splitting for hydrogen production\cite{chen2010semiconductor, corredor2019comprehensive}, \ce{CO2} reduction\cite{zhang2023coupling, li2019selective}, and pollutant degradation\cite{annadurai2023s}.
It also facilitates important processes in organic synthesis and biomedical chemistry under mild conditions. Central to these reactions lies the efficient generation, separation, and transport of photoexcited charge carriers to reactive surface sites, where they participate in redox reactions. However, the intrinsically high propensity for photogenerated electron-hole recombination severely limits solar-to-chemical conversion efficiency.\cite{zhang2025charge} Therefore, developing strategies that allow precise control over carrier migration and spatial separation is essential to suppress recombination and enhance photocatalytic performance. 

Internal electric fields, typically introduced through surface modification and interface design strategies, such as co-catalyst loading\cite{qi2020switching}, phase junctions\cite{gao2017directly}, heterojunctions\cite{fu2019ultrathin, xu2020s}, and facet junctions\cite{zhang2024crystal, li2013spatial, tada2006all}, have proven effective in promoting directional charge separation in photocatalysts. By creating interfacial potential gradients, these approaches drive photogenerated electrons and holes toward spatially distinct oxidative and reductive sites, thereby suppressing recombination and enhancing quantum efficiency. These advances underscore the pivotal role of potential landscapes in modulating carrier dynamics and offer foundational insights for next-generation photocatalyst design. 

Compared to step-like potentials created by discrete junctions, symmetry-defined periodic potentials represent a fundamentally different route to potential engineering by introducing long-range and programmable modulations in the energy landscape. These potentials can be realized through diverse platforms such as moir\'e superlattices\cite{bistritzer2011moire}, patterned electrostatic gates\cite{sun2024signature, li2021anisotropic}, or periodic strain\cite{wan2023topological}. Unlike conventional approaches, periodic potentials do not rely on chemically distinct interfaces but instead reshape the electronic structure continuously across the system. This suggests a new strategy for catalyst design. The existing literature primarily focuses on the effect of periodic modulations realized through moir\'e patterns on chemical reactivity\cite{yu2022tunable,hsieh2023domain,jiang2019mos2,zhang2024moire,li2023moire,ding2016stacking}; these studies has been restricted to system-specific and mediated by modifications of transport processes or local chemical environments, but rarely address how such potential landscapes bring wavefunction-level control of photocatalytic functionality.

%: use long-wavelength electronic structure engineering to control where photocarriers accumulate and where redox events are most likely to occur.

In this work, we establish a general mechanism in which  periodic potentials induce real-space electron-hole separation while leaving the local chemical landscape only weakly perturbed. When such a potential is applied to a two-dimensional semiconductor, it reconstructs the band structure, forms minibands, and localizes electronic wavefunctions in a symmetry-controlled manner. As a result, photoexcited electrons and holes are driven into spatially distinct regions, suppressing recombination without requiring chemical modification of the active layer. Using a minimal continuum model combined with first-principles parameters, we demonstrate this mechanism in monolayer InSe. We show that experimentally accessible moir\'e superlattices, such as those generated by twisted hBN, produce miniband formation, band-gap renormalization, and tunable carrier separation. Furthermore, by analyzing commensurate BN/InSe local registries, we establish that the moir\'e control layer transfers a measurable electrostatic modulation to the active layer, providing a microscopic bridge between continuum periodic potentials and the local surface environment. The central message is therefore not simply that moir\'e fields shift reaction descriptors, but that they offer a new and non-invasive way to build photocatalytic function by programming carrier separation in real space.

%\section{Results and Discussion}

\begin{figure}[t]
    \centering
    {\setlength{\fboxsep}{0pt}\colorbox{white}{\includegraphics[width=\columnwidth]{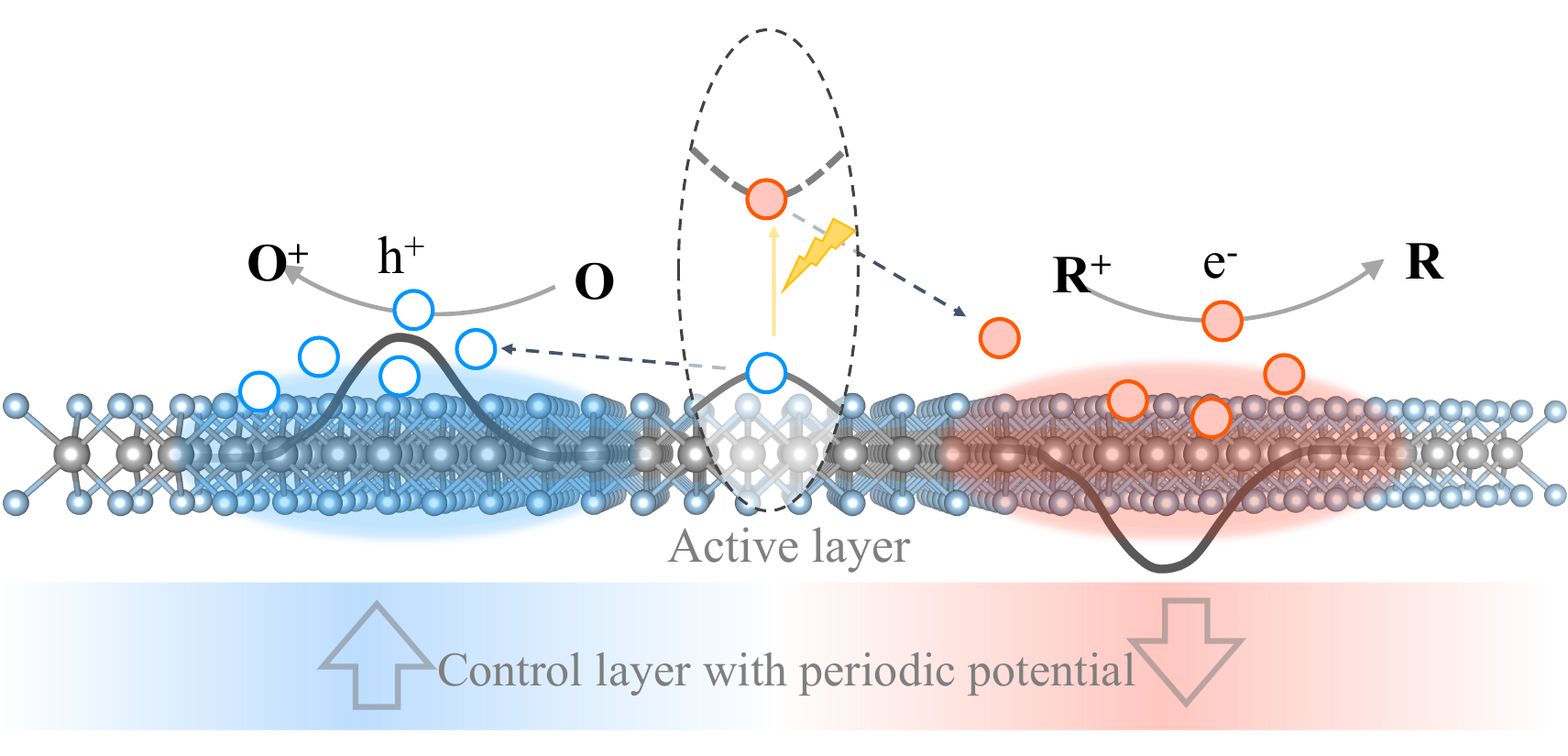}}}
    \caption{Schematic of symmetry-defined periodic-potential engineering in a two-dimensional photocatalyst. A spatially periodic potential induced by a remote patterned layer, moir\'e substrate, or strain field modulates the energy landscape of the active layer. Upon illumination, photoexcited electrons and holes are driven toward potential extrema in spatially distinct regions, enabling site-selective redox reactions and programmable catalytic functionality.}
    \label{Figure1}
\end{figure}

The proposed setup is illustrated in Fig.~\ref{Figure1}. A 2D photocatalyst layer is placed in proximity to a control layer that generates a symmetry-defined periodic potential. This potential reshapes the band structure of the active layer, forming minibands and enabling programmable control over carrier distribution. Upon illumination, the potential acts on photoexcited carriers, driving electrons and holes into spatially distinct domains and facilitating site-selective redox reactions. Such electrostatic superlattice potentials have been experimentally realized in several platforms, including moir\'e heterostructures\cite{gu2024remote, zhang2024engineering, he2024dynamically, forsythe2018band, yasuda2021stacking, wang2024band} and patterned electrostatic gates\cite{sun2024signature, li2021anisotropic}, and have also been theoretically proposed using periodic strain\cite{wan2023topological}.
To capture the physics of periodic-potential-modulated 2D photocatalysts, we adopt a minimal continuum model in which the active layer is described by a two-band $k \cdot p$ Hamiltonian, and the external modulation is introduced via a symmetry-defined electrostatic superlattice potential. For a large class of 2D semiconductor photocatalysts, the low-energy electronic structure is modeled by an effective two-band $k\cdot p$ Hamiltonian\cite{fu2007topological, bernevig2006quantum, tan2024designing}:

\begin{equation}
H_0^\tau(\mathbf{k}) =
\begin{pmatrix}
\alpha_1 |\mathbf{k}|^2 + \frac{\delta}{2} & v(\tau k_x - i k_y) \\
v(\tau k_x + i k_y) & -\alpha_2 |\mathbf{k}|^2 - \frac{\delta}{2}
\end{pmatrix}
\end{equation}
\noindent where $\mathbf{k} = (k_x, k_y)$ is the crystal momentum, $\alpha_1$ and $\alpha_2$ are quadratic dispersion parameters for the two bands, $\delta$ controls the intrinsic band gap, $v$ describes the interband coupling strength, and $\tau = \pm 1$ denotes a generalized valley index. This Hamiltonian captures both quadratic band dispersion and linear interband hybridization essential for optical transitions. All parameters ($\alpha_1$, $\alpha_2$, $v$, and $\delta$) are material- and thickness-dependent and can be extracted from first-principles DFT calculations, as detailed in the Supporting Information. To introduce periodic modulation, we consider a $C_3$-symmetric electrostatic potential, inspired by recent advances in moir\'e superlattice engineering in twisted or lattice-mismatched van der Waals heterostructures.\cite{bistritzer2011moire, angeli2021gamma, tran2019evidence} These systems naturally generate long-range periodic fields with well-defined symmetry and can be modeled as:
\begin{equation}
V(\mathbf{r}) = 2V_0 \sum_{n=1}^{3} \cos\left(\mathbf{g}_n \cdot \mathbf{r} + \phi\right)
\end{equation}
where the reciprocal lattice vectors are defined as $\mathbf{g}_n = g \left[-\cos\left(\frac{2\pi n}{3}\right), \sin\left(\frac{2\pi n}{3}\right)\right]$ with $g = 4\pi / \sqrt{3}a_M$, and $a_M$ is the real-space period of the superlattice. The strength and shape of the potential are controlled by $V_0$ and the phase parameter $\phi$. We fix $\phi = \pi$, which gives a superlattice potential with minima forming a triangular lattice and maxima forming a honeycomb lattice, an electrostatic profile commonly realized in moir\'e systems.\cite{angeli2021gamma, tran2019evidence}

\begin{figure*}[t]
    \centering
    {\setlength{\fboxsep}{0pt}\colorbox{white}{\includegraphics[width=0.9\textwidth]{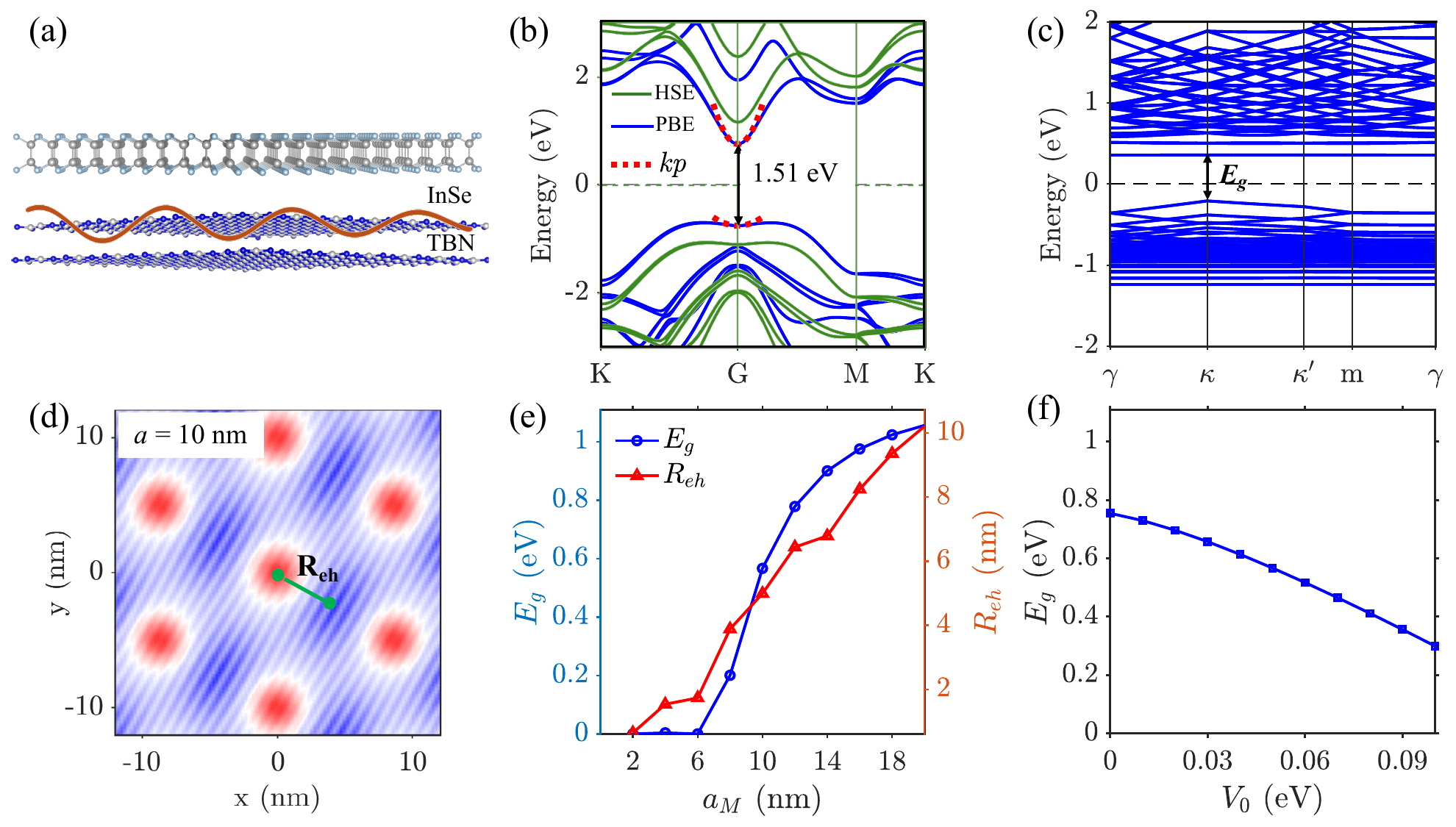}}}
    \caption{Effect of modulated periodic potentials on band structure and real-space carrier separation in monolayer InSe.
(a) Schematic for using twisted BN as a moir\'e polar substrate. InSe monolayer sits on top of twisted bilayer hBN, feeling its periodic moir\'e potential. (b) Band structure of the unmodulated two-band $k \cdot p$ model fitted with the DFT band structure at PBE level. The HSE06 band structure is also plotted for comparison. 
(c) Miniband formation and strong band reconstruction under a $C_3$-symmetric periodic potential ($V_0 = 50~meV$, $a_M = 10~nm$, $\phi = \pi$). (d) Real-space distribution of photoexcited carriers under periodic modulation, with electrons (red) and holes (blue) localized at distinct regions. The separation distance $R_{e\text{--}h}$ is defined as the distance between their respective charge density centers. (e) The band gap ($E_g$) and $R_{e\text{--}h}$ of the engineered two-dimensional superlattice as a function of superlattice period $a_M$. (f) $E_g$ as a function of superlattice potential strength $V_0$.}
    \label{Figure2}
\end{figure*}

Monolayer InSe, exfoliated from bulk $\gamma$-phase crystals by mechanical\cite{lei2014evolution} or vapor-phase methods\cite{ho2016thickness}, exhibits excellent ambient stability and high carrier mobility\cite{bandurin2017high}, making it a promising candidate for ultrathin optoelectronic and photocatalytic applications. However, its indirect band gap and weak internal electric fields limit the efficiency of carrier generation and redox activity. The symmetry-defined periodic potential strategy introduced here offers a general and non-invasive route to tailor its electronic structure, enhance charge separation, and potentially unlock its latent photocatalytic functionality without altering the intrinsic lattice. A material-specific $k \cdot p$ description of monolayer InSe was therefore constructed from first-principles calculations performed with VASP and PAW potentials.\cite{hafner2008ab, kresse1996efficiency, kresse1996efficient, kresse1999ultrasoft} Fig.~\ref{Figure2}b shows that monolayer InSe has an indirect bandgap and a characteristic sombrero-shaped valence-band edge. The calculated bandgaps are 2.26 eV at the HSE06 level and 1.46 eV at the PBE level, consistent with earlier work.\cite{he2019improvement, zhuang2013single, peng2017computational} Because the energy difference between the valence-band edge and the top of the valence band at $\Gamma$ is small (45/50 meV at the HSE06/PBE levels), an effective model centered at $\Gamma$ captures the relevant low-energy physics. The fitted four-band Hamiltonian reproduces the first-principles band structure well; technical details and parameters are deferred to the Supporting Information.

Recent experimental work has demonstrated that few-layer InSe encapsulated by hexagonal boron nitride (hBN) exhibits exceptional structural quality and ambient stability, along with high carrier mobilities exceeding $10^3$-$10^4$\,cm$^2$\,V$^{-1}$\,s$^{-1}$ at room and cryogenic temperatures.\cite{bandurin2017high} Owing to the atomically flat and lattice-matched nature of hBN, a twisted bilayer hBN can generate a long-wavelength moir\'e superlattice as the control layer for the InSe, providing an experimentally viable platform to implement the symmetry-defined periodic potentials proposed in this work (Fig.~\ref{Figure2}a).
Recent experimental realizations of twisted bilayer hBN have demonstrated tunable periodic potentials with amplitudes from 10 to 100 meV and wavelengths ranging from 5 to 50 nm,\cite{wang2025moire} establishing the practical feasibility of such long-range modulations.
To ensure the applicability of our continuum framework under such conditions, we assess the validity of the underlying $k \cdot p$ expansion. For InSe, the effective models remain accurate within a momentum window of $|\mathbf{k}| \lesssim 0.2$~\AA$^{-1}$. This sets a lower bound on the real-space period $a_m \gtrsim 3.6$ nm. Thus, the experimentally accessible moir\'e wavelengths fall well within the validity regime of our model.

Fig.~\ref{Figure2}c illustrates the reconstructed band structure of monolayer InSe under a realistic periodic potential ($V_0 = 50$\,meV, $a_M = 10$\,nm), showing clear miniband formation and strong hybridization near the band edges. The central physical consequence is seen in real space: the periodic landscape drives electrons and holes into distinct regions of the moir\'e unit cell (Fig.~\ref{Figure2}d). This spatial carrier separation is the main functionality generated by the periodic potential and is the key ingredient that makes the proposal relevant for photocatalysis. As the superlattice period $a_M$ increases, the bandgap gradually recovers, while the electron-hole separation length $R_{e\text{--}h}$ grows more rapidly (Fig.~\ref{Figure2}e). This trade-off highlights a design principle for optimizing carrier separation without relying on chemical modification of the active layer.
In addition, increasing the potential strength $V_0$ reduces the bandgap, providing an additional tuning knob for engineering carrier dynamics and band structure (Fig.~\ref{Figure2}f).

\begin{figure}[t]
    \centering
    {\setlength{\fboxsep}{0pt}\colorbox{white}{\includegraphics[width=0.9\columnwidth]{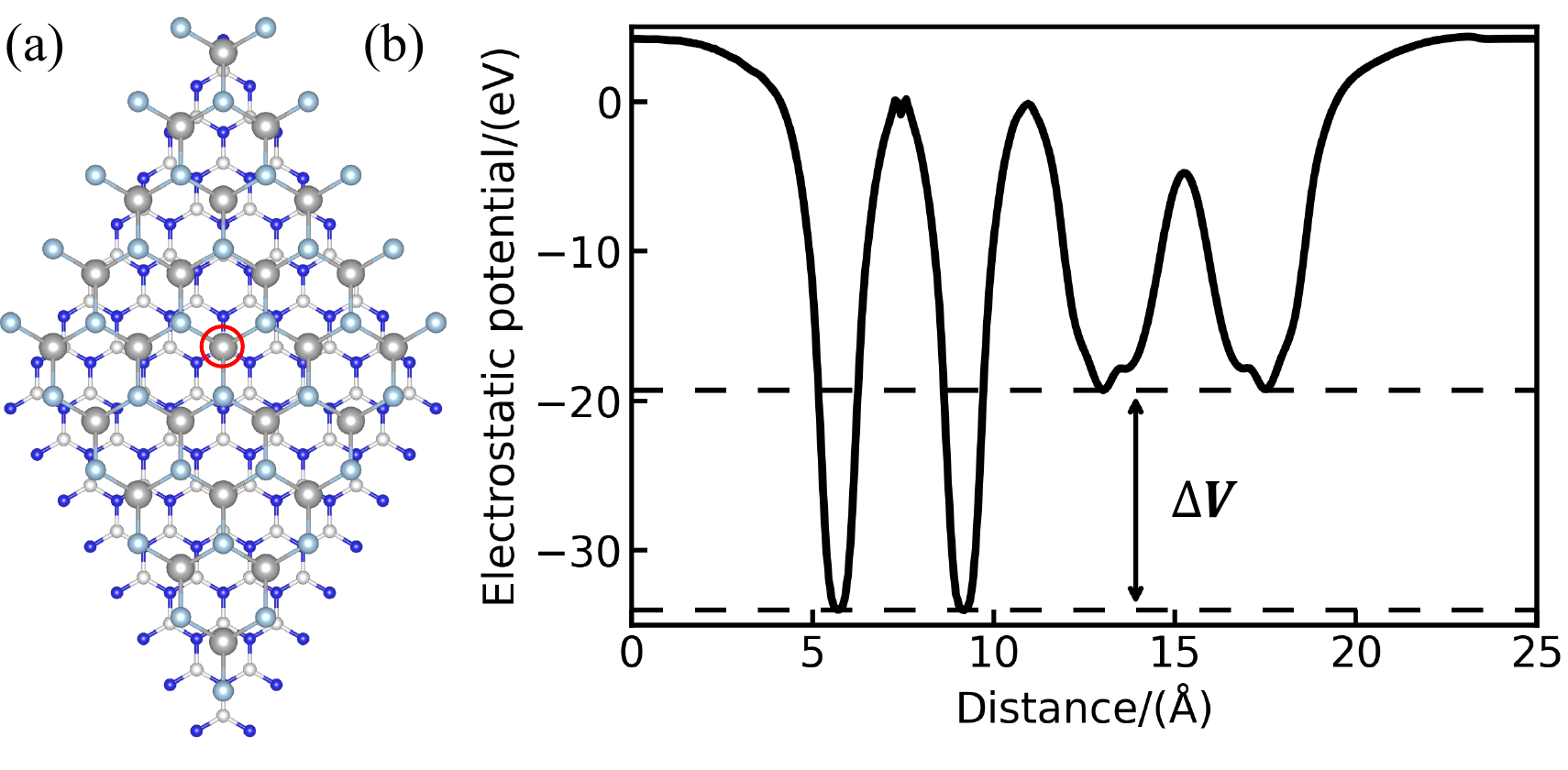}}}
    \caption{\textbf{Registry-dependent electrostatic potential transfer in BN/InSe heterostructures.} Electrostatic potentials extracted from representative BN/InSe local registries quantify how the BN environment imprints a local modulation onto the adjacent InSe layer. The layer-resolved offset is defined as $\Delta V = |V_{\min}^{\mathrm{InSe}} - V_{\min}^{\mathrm{BN}}|$, where the minima are obtained from the planar-averaged electrostatic potential $V(z)$ within the BN and InSe regions. The registry-induced spread $\delta V_{\mathrm{reg}} = \max(\Delta V) - \min(\Delta V)$ isolates the transferable component of the electrostatic modulation and provides the microscopic bridge to the local-field picture used below.}
    \label{Figure3}
\end{figure}

%\subsection{Electrostatic Potential Transfer from BN to InSe}
An essential question is whether a realistic moir\'e platform can transfer a measurable electrostatic modulation to the active InSe layer. To address this point, we analyze BN/InSe local registries that serve as commensurate approximants to the moir\'e environment generated by twisted BN. We consider nine representative stackings spanning the AA, AB, and BA families, with In located above B, N, and hollow sites. For each registry, the planar-averaged electrostatic potential $V(z)$ is extracted from first-principles calculations, and the layer-resolved offset is defined as $\Delta V = |V_{\min}^{\mathrm{InSe}} - V_{\min}^{\mathrm{BN}}|$. Because $\Delta V$ contains a large common background contribution, the more relevant quantity is the registry-induced spread $\delta V_{\mathrm{reg}} = \max(\Delta V) - \min(\Delta V)$. From the nine local registries, we obtain $\delta V_{\mathrm{reg}} \approx 0.041$\,eV, demonstrating that twisted BN can imprint a finite and spatially varying electrostatic landscape onto the adjacent InSe layer. This establishes the missing microscopic bridge between the continuum periodic-potential picture and the local surface response discussed below.
\begin{figure}[t]
    \centering
    {\setlength{\fboxsep}{0pt}\colorbox{white}{\includegraphics[width=\columnwidth]{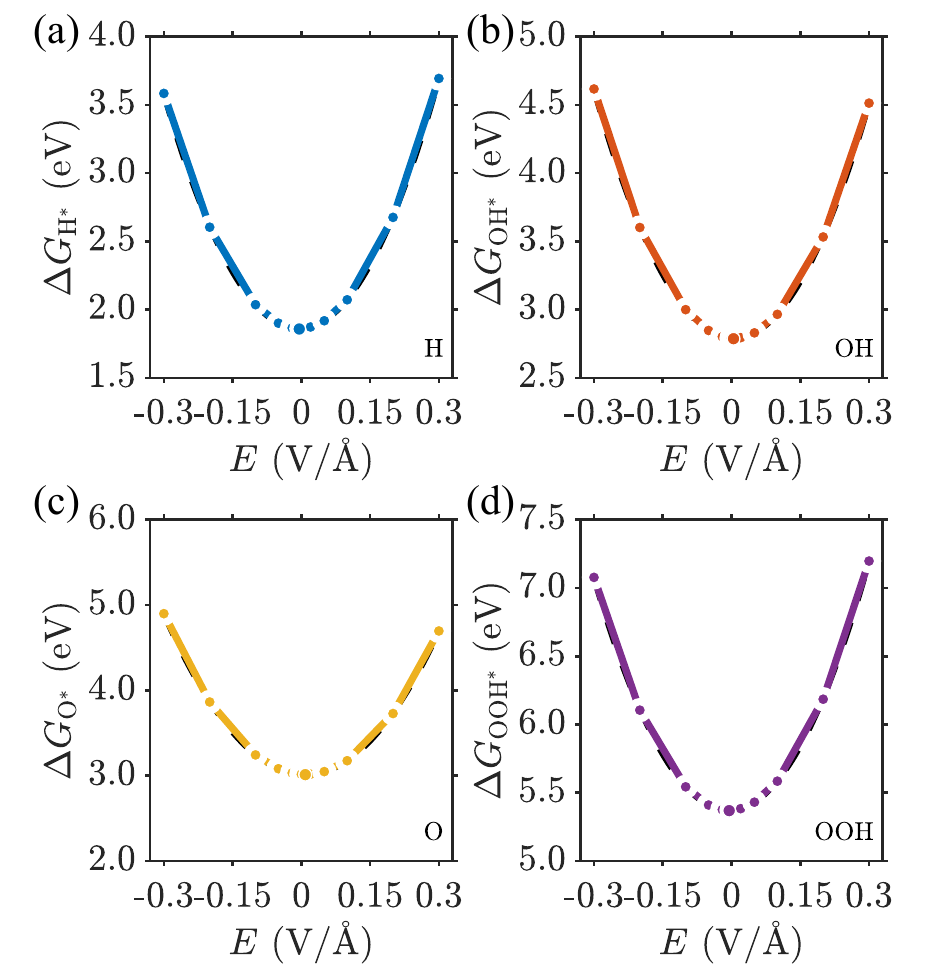}}}
    \caption{Field-dependent Gibbs free-energy descriptors for catalytic intermediates on monolayer InSe. Field dependence of the adsorption free energies of H$^*$, OH$^*$, O$^*$, and OOH$^*$ under an out-of-plane electric field. All descriptors show a systematic, approximately parabolic variation with field strength and remain closest to their minimum values near zero field, indicating that the applied field acts primarily as a tuning parameter for the local adsorption landscape within the scanned window.}
    \label{Figure4}
\end{figure}

%\subsection{Weak Surface Response to the Moir\'e Potential}
To test whether the transferred electrostatic landscape also strongly rewrites the surface chemistry, we evaluate standard HER and OER adsorption free energies on monolayer InSe under out-of-plane fields that mimic different local moir\'e environments. Here the adsorption free energy simply measures how favorable a reaction intermediate is on the surface. The most favorable adsorption site is the top-Se site for H$^*$, while the preferred adsorption configurations for OH$^*$, O$^*$, and OOH$^*$ are the top-In, top-Se, and hollow sites, respectively, as summarized in Fig.~S3. Figure~\ref{Figure4} shows that all four descriptors vary only weakly and approximately parabolically with the applied field. This behavior is consistent with a Stark-response picture, in which the local electrostatic environment acts only as a perturbation to the adsorption complex rather than as a strong driver of chemical reconstruction.

Using the BN-InSe distance of 3.39~\AA, we estimate an effective field of $E_{\mathrm{eff}} \sim \delta V_{\mathrm{reg}}/d \approx 0.012$~V/\AA. This places the system in a weak-coupling regime: the moir\'e potential is strong enough to reorganize carrier distribution, but too small to produce large changes in adsorption trends. The detailed HER and OER free-energy diagrams in Fig. S4 therefore play a supporting role. They show that pristine InSe is not transformed into a strongly active catalyst by the moir\'e field alone. Taken together, the results provide a realistic route toward decoupled design of charge separation and surface reactivity for future photocatalyst designs, suggesting that the periodic potentials should be viewed primarily as a tool for programmable charge separation, to be combined with chemically favorable surfaces.

%\section{Discussion}
Photoexcited carriers in two-dimensional semiconductors frequently form bound excitons because of reduced dielectric screening and enhanced Coulomb interactions. Recent studies report a relatively small exciton binding energy in monolayer InSe ($\sim 0.2$ eV)\cite{ceferino2020crossover, peng2017computational}, which is nonetheless non-negligible compared with the typical amplitude of the periodic potentials considered here. Notably, even in the excitonic regime, moir\'e-modulated systems have been shown to exhibit substantial real-space separation between the electron and hole, particularly when the moir\'e period exceeds the exciton radius. These findings suggest that periodic potentials can reshape the spatial structure of excitons and suppress recombination even when exciton binding remains significant. Although our present study does not explicitly include excitonic effects, the single-particle framework remains appropriate for capturing the primary mechanism of real-space carrier separation under moir\'e superlattices. A more complete treatment of exciton formation, binding modulation, and dynamics in such engineered potential landscapes is left for future work.

Conceptually, our results shift the role of moir\'e-scale electrostatic modulation from an electronic-structure effect to a functional design principle for photocatalysis. In the regime identified here, periodic potentials strongly redistribute electronic wavefunctions and drive spatial separation of electrons and holes, without substantially altering the intrinsic surface chemistry. This suggests a general strategy: rather than relying on a single material to simultaneously achieve charge separation and catalytic activity, one can combine moir\'e-engineered carrier separation with materials that are already chemically active. The coupling between the programmed electrostatic landscape and interfacial reactions can be further enhanced by introducing dipoles, polar adsorbates, or intrinsically polar layers. In this sense, monolayer InSe serves as a proof-of-principle platform for a broader class of photocatalytic architectures based on programmable long-wavelength potentials.

%\section{Conclusions}
In this work, we propose periodic potentials as a new strategy for photocatalysis in two-dimensional semiconductors. Within a minimal continuum description combined with first-principles parameters for monolayer InSe, we show that such potentials reconstruct the band structure, induce miniband formation, and generate robust real-space separation of photoexcited carriers. The accompanying moir\'e-imprinted electrostatic modulation is measurable but remains in a weak-coupling regime, so the local surface chemistry is only modestly affected. The resulting picture is therefore simple and general: long-wavelength periodic potentials can be used to engineer where electrons and holes go after photoexcitation, without requiring major chemical redesign of the active surface. This identifies a new path for photocatalyst design based on programmable carrier separation and opens a broader route toward moir\'e-enabled photochemistry, light-driven interfacial control, and catalytic architectures built from electronically engineered two-dimensional materials.

\textit{Acknowledgments}-The authors acknowledge useful discussions with Yiyang Jiang, Changjiang Yi, Arpit Arora, Tixuan Tan, and Zexun Lin. Q.Y. acknowledges the funding from the Otto Hahn Award from the Max Planck Society with funding No. 85931. 

\bibliographystyle{apsrev4-2}
\bibliography{reference}

\end{document}